\documentclass[osajnl,twocolumn,showpacs,superscriptaddress,10pt,groupedaddress]{revtex4-1}

\usepackage{amsmath,amssymb}
\usepackage[english]{babel}
\usepackage{graphicx}
\usepackage{braket}
\usepackage{bbm}
\usepackage{float}
\usepackage{dcolumn}
\usepackage{color}

\renewcommand{\vec}{\mathbf}

\begin{document}

\title{Feasibility of UV lasing without inversion in mercury vapor}
\author{Martin R. Sturm}
\email[e-mail: ]{martin.sturm@physik.tu-darmstadt.de}
\author{Benjamin Rein}
\author{Thomas Walther}
\author{Reinhold Walser}
\affiliation{Institut f\"ur Angewandte Physik, Technische Universit\"at Darmstadt, Hochschulstra\ss e 4A, Darmstadt D-64289, Germany}

\begin{abstract}
We investigate the feasibility of UV lasing without inversion at a wavelength of $253.7$ nm utilizing interacting dark resonances in mercury vapor. Our theoretical analysis starts with radiation damped optical Bloch equations for all relevant 13 atomic levels. These master equations are generalized by considering technical phase noise of the driving lasers. From the Doppler broadened complex susceptibility we obtain the stationary output power from semiclassical laser theory. The finite overlap of the driving Gaussian laser beams defines an ellipsoidal inhomogeneous gain distribution. Therefore, we evaluate the intra-cavity field inside a ring laser self-consistently with Fourier optics. This analysis confirms the feasibility of UV lasing and reveals its dependence on experimental parameters.
\end{abstract}


\maketitle

\newcommand{\eq}[1]{Eq.~\eqref{#1}}


\section{Introduction}
Developing powerful, coherent light sources ranging from UV to X-ray is a major quest in laser development with relevant applications from spectroscopy, lithography to material science. Conventional lasing requires population inversion, which becomes increasingly difficult for shorter wavelengths since the threshold pumping power scales with the laser frequency $\omega^4$ to $\omega^6$. In the UV regime lasing without inversion (LWI) is a possible pathway to overcome this problem \cite{Kocharovskaya1988, Harris1989, Scully1989, Kocharovskaya1992, Scully1994, Lukin1996, Mompart2000}.

To date, several experiments \cite{Nottelmann1993,Fry1993,Zibrov1995,Kleinfeld1996} have been conducted showing that inversionless lasing is in fact feasible. However, the lasing wavelengths were not significantly shorter than the driving fields' wavelengths. Despite all commitment, a laser based on the LWI concept operating in the UV regime is yet to be built. The large majority of existing UV lasers are based on nonlinear harmonic frequency generation. Developing an alternative to this technique using LWI might allow for new applications.

Doppler broadening is a major obstacle in UV lasing without inversion when driving frequencies are strongly disparate \cite{Lukin1996}. One path to circumvent this problem is transient lasing without inversion \cite{Kilin2008, Svidzinsky2013, Yuan2014}. However, it is limited to pulsed lasing. Another path allowing for Doppler-free cw LWI has been proposed by Fry {\sl et al.} \cite{Fry1999}. It is based on the concept of interacting dark resonances \cite{Lukin1999}. The proposed experiment allows for lasing on the $6^3P_1 \leftrightarrow 6^1S_0$ transition in mercury at a wavelength of $253.7$ nm. This idea can also be applied to similar schemes for example in mercury and krypton \cite{Takeoka1998} at wavelengths of $185$ nm and $116.5$ nm respectively.

In this paper, we provide a realistic three-dimensional theoretical analysis of the experiment proposed by Fry {\sl et al.}. The article is structured as follows. In Secs.~\ref{darkRes} and \ref{radDamp} the basic LWI scheme is introduced and applied to the realistic 13-level scheme of mercury. Sec.~\ref{doppler} introduces the Doppler-free three-photon resonance \cite{Fry1999}, which shields the linear gain coefficient from inhomogeneous line broadening. Further broadening effects are considered in Sec.~\ref{otherBroad}. We consider technical phase noise of the driving fields and assess its effect on laser gain in Sec.~\ref{noise}.  In Sec.~\ref{lasertheory}, the stationary laser power is calculated using self-consistent semiclassical laser theory. In the concluding Sec.~\ref{resonatorModes}, we use the linear gain coefficient of the spatially inhomogeneous mercury vapor and evaluate the intracavity field modes of a four-mirror ring laser resonator self-consistently within Fourier optics.


\section{Interacting dark resonances}
\label{darkRes}
\begin{figure}
	\centering
	\includegraphics[scale=1]{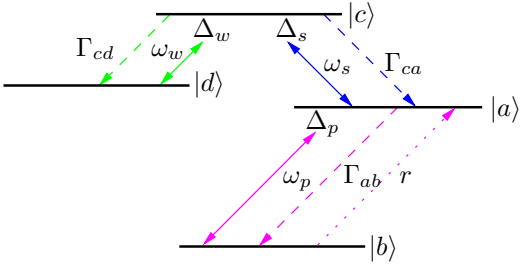}
	\caption{Basic four-level coupling scheme for LWI. The solid arrows represent coherent coupling, the dashed arrows incoherent decay and the dotted arrow incoherent pumping. Here $a \leftrightarrow b$ represents the probe transition, while $c\leftrightarrow a$ and $c \leftrightarrow d$ are the coherent driving transitions.}
	\label{fig:fourlscheme}
\end{figure}

We implement LWI in a four-level scheme, as shown in Fig.~\ref{fig:fourlscheme}. In such a scheme with three allowed dipole transitions driven by a strong and a weak external electric field $\vec{E}_s$ and $\vec{E}_w$ respectively and a probe field $\vec{E}_p$, one finds interacting dark resonances.

For each of these fields, $j=s,w,p$, the positive frequency components are given by
\begin{align}
\vec{E}_j^{(+)} (\vec{r},t) &= \mathcal{E}_j (\vec{r},t) \boldsymbol{\epsilon}_j \exp \left( -i \omega_j t \right),\label{eqn:electricfields}
\end{align}
with the angular frequencies $\omega_j$, polarization vectors $\boldsymbol{\epsilon}_j$, and slowly varying amplitudes $\mathcal{E}_j$. For each of the three dipole transitions we obtain the corresponding Rabi frequencies $\Omega_p = \vec{d}_{ab} \cdot \boldsymbol{\epsilon}_p \mathcal{E}_p / \hbar$, $\Omega_s = \vec{d}_{ca} \cdot \boldsymbol{\epsilon}_s \mathcal{E}_s / \hbar$, and $\Omega_w = \vec{d}_{cd} \cdot \boldsymbol{\epsilon}_w \mathcal{E}_w / \hbar$ with the dipole matrix elements of the respective transitions $\vec{d}_{ab}$, $\vec{d}_{ca}$, and $\vec{d}_{cd}$.

Within the dipole and rotating wave approximation \cite{Schleich2001}, one finds for the Hamiltonian matrix
\begin{equation}
	H= -\hbar \begin{pmatrix} 0 & \Omega_p^* & \Omega_s & 0 \\
	\Omega_p & -\Delta_p & 0 & 0 \\
	\Omega_s^* & 0 & \Delta_s & \Omega_w^* \\
	0 & 0 & \Omega_w & \Delta_s - \Delta_w
\end{pmatrix},
\label{eqn:hamiltonianmatrix}
\end{equation}
where the matrix elements are sorted in the order of the basis $\left\{ \ket{a},\ket{b},\ket{c},\ket{d} \right\}$. The detunings are defined as $\Delta_p = \omega_p - (\omega_a - \omega_b)$, $\Delta_s = \omega_s - (\omega_c - \omega_a)$, and $\Delta_w = \omega_w - (\omega_c - \omega_d)$ with $\hbar \omega_j$ being the energy of the respective atomic state.

The origin of lasing without inversion can be understood best in the dressed state picture. There, one finds for the eigenstates and energies of the Hamiltonian
\begin{align}
	\ket{0} &=  \ket{d} - \frac{\Omega_w^*}{\Omega_s^*} \ket{a},& E_0 &= 0, \\
	\ket{\pm} &= \frac{1}{\sqrt{2}} \left( \ket{a} \mp \frac{\vert \Omega_s \vert}{\Omega_s} \ket{c} +\frac{\Omega_w}{\Omega_s} \ket{d}\right),& E_\pm &= \pm \hbar \vert \Omega_s \vert,
\end{align}
assuming vanishing $\Omega_p$ and first order contributions in $\Omega_w$. For the sake of simplicity, resonant coupling fields were chosen. For vanishing $\Omega_w$, the states $\ket{+}$ and $\ket{-}$ correspond to the well known Autler-Townes doublet of the three-level ladder system $\ket{b}$, $\ket{a}$, and $\ket{c}$. Hence, probing the transition $a\leftrightarrow b$, one observes the Autler-Townes splitting in the absorption spectrum. For finite values of $\Omega_w$, the state $\ket{0}$, originally corresponding to the bare state $\ket{d}$, contains an admixture of $\ket{a}$ and by this means couples to $\ket{b}$. This transition corresponds to the three-photon transition $d \leftrightarrow b$ in the bare state picture and is responsible for a sharp absorption feature that can be used for LWI as we will see in the following section.

\section{Radiation damped optical Bloch equations}
\label{radDamp}
As atoms are embedded in an open system they experience radiation damping, which is described by means of optical Bloch equations
\begin{equation}
	\partial_t \hat{\rho} = ( \mathcal{L}_c + \mathcal{L}_i) \hat{\rho}
	\label{eqn:Bloch}
\end{equation}
for the reduced density operator $\hat{\rho}$ of the atomic system. Within the Born-Markov approximation \cite{Barnett1997,Louisell1973}, one finds for the coherent evolution the Liouvillian
\begin{equation}
\mathcal{L}_c \hat{\rho} = -\frac{i}{\hbar} \left[ \hat{H} ,\hat{\rho} \right],
\label{eqn:cohLiou}
\end{equation}
which is the free Hamiltonian evolution of a multi-level atom in presence of coherent laser radiation, for example \eq{eqn:hamiltonianmatrix}. Irreversible radiation damping is represented by the incoherent Liouvillian
\begin{equation}
\begin{split}
\mathcal{L}_i \hat{\rho} = &\sum\limits_{k \in \mathcal{D}}
	\Gamma_k \frac{n_k+1}{2} \left( 2 \hat{s}_k \hat{\rho} \hat{s}_k^\dagger - \hat{s}_k^\dagger \hat{s}_k \hat{\rho}
	-\hat{\rho} \hat{s}_k^\dagger \hat{s}_k\right) \\ 
	&+ \sum\limits_{k \in \mathcal{D}} \Gamma_k \frac{n_k}{2} \left( 2
	\hat{s}_k^\dagger \hat{\rho} \hat{s}_k - \hat{s}_k \hat{s}_k^\dagger \hat{\rho} -\hat{\rho} \hat{s}_k
	\hat{s}_k^\dagger\right).
\end{split}
\end{equation}
The sum extends over all allowed transitions $\mathcal{D} = \left\{ ab, ca, cd \right\}$ and the corresponding natural decay rates $\Gamma_{ab}$, $\Gamma_{ca}$ and $\Gamma_{cd}$. The mean photon number of the transition $k$ is labeled by $n_k$ and corresponds to the lowering operator $\hat{s}_k$. For the scheme shown in Fig.~\ref{fig:fourlscheme}, they are defined as $\hat{s}_{ab} = \ket{b}\bra{a}$, $\hat{s}_{ca} = \ket{a}\bra{c}$, and $\hat{s}_{cd} = \ket{d}\bra{c}$. Since the thermal population of optical modes is negligible in the proposed experiment, we choose $n_{ca}=n_{cd}=0$. However, in order to model an incoherent, bidirectional pump on the lasing transition $a \leftrightarrow b$, we set the photon number $n_{ab} = r/\Gamma_{ab}$ proportional to the pump rate $r$.

The polarization density of the gas is
\begin{equation}
\vec{P} = \mathcal{N} \, \text{Tr} \left\{ \hat{\vec{d}} \hat{\rho} \right\},
\end{equation}
with the atomic density $\mathcal{N}$ and the dipole operator $\hat{\vec{d}}$. The part of the polarization density associated to the probe transition is spectrally well separated from the other parts and will be called $\vec{P}_p$.

In the mercury vapor cell the atomic density is calculated for a given temperature $T$ using the ideal gas law and the vapor pressure of mercury \cite{Roth1990,Villwock2010}. If not specified otherwise, a temperature of $T=300$ K is used for calculations in this paper, resulting in an atomic density of $\mathcal{N}=9.2 \times 10^{13}$ cm$^{-3}$.

\begin{figure}
	\includegraphics[scale=1]{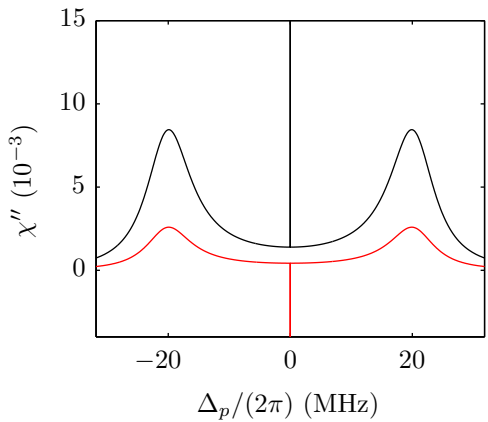}
	\caption{Absorption $\chi''(\Delta_p)$ on the lasing transition versus detuning $\Delta_p$ with $\Omega_s = 2\pi \times 20.7$ MHz, $\Omega_w = 2\pi \times 0.3$ MHz, and $\Delta_s = \Delta_w = 0$. In the absence of incoherent pumping $r=0$, we obtain pure absorption (black line), but in the presence of an incoherent pump $r=5$ kHz, one finds gain (red line). The characteristic features of the plot are the Autler-Townes doublet at $\pm \Omega_s$, as well as the extremely sharp LWI resonance ($\sim 1$ kHz) in the center. For representation purposes, we have truncated the central peaks.}
	\label{fig:fourls}
\end{figure}

Assuming linear response \cite{Kubo1957} and an isotropic medium, the polarization density on the probe transition is proportional to the applied field
\begin{equation}
\vec{P}^{(+)}_p  = \epsilon_0 \chi^{(1)} \vec{E}^{(+)}_p.
\end{equation}
defining the linear complex susceptibility as
\begin{equation}
\chi^{(1)} = \chi^\prime + i \chi^{\prime\prime} = \frac{\vert d_{ab} \vert^2 \mathcal{N}}{\epsilon_0 \hbar \Omega_p} \rho_{ab}.
\end{equation}
The susceptibility's real part $\chi^\prime$ accounts for dispersion, while its imaginary part $\chi^{\prime\prime}$ describes absorption/gain. For $\chi^{\prime\prime}>0$ the probe field is attenuated (absorption), while for $\chi^{\prime\prime}<0$ it is amplified (gain). To find the linear absorption spectrum $\chi^{\prime\prime}(\Delta_p)$, we solved \eq{eqn:Bloch} in the stationary limit. Fig.~\ref{fig:fourls} shows the resulting spectra.
The absorption spectra show peaks at $\Delta_p = \pm \vert \Omega_s \vert$, corresponding to the Autler-Townes splitting as expected. At the three-photon resonance $\Delta_p = \Delta_w -\Delta_s$ in the center of the spectra a very sharp peak occurs that can be explained by the interaction of the dark resonances $\ket{0}$ and $\ket{b}$ \cite{Lukin1999,Fry1999}.

These results reveal that even a small amount of incoherent pumping can invert the sharp absorption peak into a gain dip and lead to lasing on the probe transition. The fact that only the central peak is inverted, while the rest of the spectrum is merely unchanged, indicates that the population on the lasing transition is not inverted. Based on this  idea \cite{Lukin1999}, a proposal for an LWI experiment in mercury was published in \cite{Fry1999} that allows for cw lasing in the UV and VUV regime. The goal of the following sections will be to investigate the feasibility of this proposal with regard to a realistic setting. 
\subsection{Realistic coupling scheme for mercury}
\begin{figure}
	\centering
	\includegraphics[scale=1]{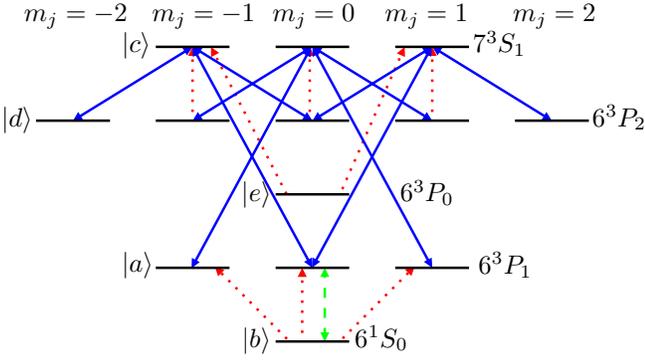}
	\caption{Coupling scheme of mercury including all relevant states for the experiment and their Zeeman structure. The optical couplings are: the coherent driving fields (solid blue), the incoherent pumping fields and the spectrally broad repump field (dotted red), the lasing transition (dashed green).}
	\label{fig:fullls}
\end{figure}

Following \citep{Fry1999}, we introduce the full coupling scheme (Fig.~\ref{fig:fullls}) for all relevant two-electron states of mercury \citep{Sobelman1991} including an additional trapping state $\ket{e}$. These 13 states can be classified by the Zeeman manifolds
\begin{align}
\mathcal{Z}_a &= \left\{ \ket{n_a=6,J_a=1,m_a = 0,\pm1} \right\}, \\
\mathcal{Z}_b &= \left\{ \ket{n_b=6,J_b=0,m_b = 0} \right\}, \\
\mathcal{Z}_c &= \left\{ \ket{n_c=7,J_c=1,m_c= 0,\pm1} \right\}, \\
\mathcal{Z}_d &= \left\{ \ket{n_d=6,J_d=2,m_d = 0,\pm1,\pm2} \right\}, \\
\mathcal{Z}_e &= \left\{ \ket{n_e=6,J_e=0,m_e =0} \right\}.
\end{align}
The atomic states are labeled as $\ket{n_i,J_i,m_i}$ with the principal quantum number $n_i$, the total angular momentum quantum number $J_i$, and the projection quantum number $m_i$.

To prevent population trapping in $\ket{e}$, a repump field
\begin{equation}
\vec{E}_r^{(+)} (\vec{r},t) = \mathcal{E}_r (\vec{r},t) \boldsymbol{\epsilon}_r \exp \left( -i \omega_r t \right),
\end{equation}
is introduced with a detuning $\Delta_r = \omega_r - (\omega_c - \omega_e)$. Further, an incoherent pump rate $r_{cd}$ on the $\Delta m = 0$ transitions between $\ket{c}$ and $\ket{d}$ prevents population trapping in $\ket{6,2,\pm1}$.

Using a convenient interaction picture, the system's Hamiltonian $\hat{H} = \hat{H}_0 + \hat{V}$ in dipole and rotating wave approximation is given by
\begin{align}
\begin{split}
\hat{H}_{0} = &\hbar\Delta_p \hat{s}_{bb} - \hbar\Delta_s \hat{s}_{cc}  + \hbar\left( \Delta_w - \Delta_s \right) \hat{s}_{dd} \\
&+ \hbar\left( \Delta_w - \Delta_r \right) \hat{s}_{ee},
\label{eqn:H0}
\end{split}\\
\begin{split}
\hat{V} = &-\hbar \sum\limits_{q = 0,\pm 1} \big(\Omega_{p}^q \hat{s}_{ab}^q + \Omega_{s}^q \hat{s}_{ca}^q \\
 &+ \Omega_{w}^q \hat{s}_{cd}^q + \Omega_{r}^q \hat{s}_{ce}^q \big)+ \text{H.c.}\,.
 \end{split}
\label{eqn:HI}
\end{align}
The projection operators are defined as
\begin{equation}
\hat{s}_{jj} = \sum\limits_{k \in \mathcal{Z}_j} \ket{k} \bra{k},
\end{equation}
while the lowering operators are given by
\begin{align}
\begin{split}
\hat{s}_{ij}^q = &\sum\limits_{\mathcal{Z}_i,\mathcal{Z}_j} (-1)^{J_i-m_i} \sqrt{2J_i+1} \\
&\times \begin{pmatrix}
J_i & 1 & J_j \\
-m_i & q & m_j
\end{pmatrix} \ket{n_j,J_j,m_j} \bra{n_i,J_i,m_i},
\end{split}
\end{align}
employing Wigner 3-j symbols \cite{Sobelman1991} and the spherical polarization vectors $\vec{e}_{q=0} = \vec{e}_z$, $\vec{e}_{q=\pm 1} = \mp (\vec{e}_x \pm i \vec{e}_y)/\sqrt{2}$. The Rabi frequencies polarization components are given by
\begin{align}
\Omega_p^q &= \vec{e}_{q}^* \cdot \boldsymbol{\epsilon}_{p}  \frac{ \langle a \Vert \hat{d} \Vert b \rangle \mathcal{E}_p}{\hbar \sqrt{2 J_a +1}},&
\Omega_s^q &= \vec{e}_{q}^* \cdot \boldsymbol{\epsilon}_{s}  \frac{ \langle c \Vert \hat{d} \Vert a \rangle \mathcal{E}_s}{\hbar \sqrt{2 J_c +1}}, \\
\Omega_w^q &= \vec{e}_{q}^* \cdot \boldsymbol{\epsilon}_{w}  \frac{ \langle c \Vert \hat{d} \Vert d \rangle \mathcal{E}_w}{\hbar \sqrt{2 J_c +1}},&
\Omega_r^q &= \vec{e}_{q}^* \cdot \boldsymbol{\epsilon}_{r}  \frac{ \langle c \Vert \hat{d} \Vert e \rangle \mathcal{E}_r}{\hbar \sqrt{2 J_c +1}},
\end{align}
with $\langle i \Vert \hat{d} \Vert j \rangle$ being the $i \leftrightarrow j$ transition's reduced dipole matrix element \cite{Sobelman1991}. Experimental values have been taken from the NIST database \cite{NISTdatabase} and are summarized in Table \ref{table}.
\begin{table}[H]
\caption{Properties of the atomic transitions: transition, wavelength, natural line width, line strengths $S_{ij}=\langle i \Vert \hat{d} \Vert j \rangle^2$ in atomic units (Bohr radius $a_0$ and elementary charge $e$).}
\begin{ruledtabular}
\begin{tabular}{cddd}
Transition & \multicolumn{1}{l}{$\lambda$ (nm)} & \multicolumn{1}{l}{$\Gamma$ (MHz)} & \multicolumn{1}{l}{$S$ ($a_0^2 e^2$)} \\
\hline
$a \leftrightarrow b$ & 253.7 & 2\pi \times 1.27 & 0.19\\
$c \leftrightarrow a$ & 435.8 & 2\pi \times 8.86 & 6.83\\ 
$c \leftrightarrow d$ & 546.1 & 2\pi \times 7.75 & 11.8\\ 
$c \leftrightarrow e$ & 404.7 & 2\pi \times 3.45 & 2.1
\end{tabular}
\end{ruledtabular}
\label{table}
\end{table}

The laser fields' polarization vectors are given by
\begin{align}
\boldsymbol{\epsilon}_p &= \vec{e}_x & \boldsymbol{\epsilon}_s &= \boldsymbol{\epsilon}_w = \boldsymbol{\epsilon}_r = \vec{e}_y. \label{eqn:pol}
\end{align}
The system's Bloch equations generalize \eq{eqn:Bloch} by summing over all polarizations and allowed transitions.

\section{Doppler broadening}
\label{doppler}
\begin{figure}
	\centering
	\includegraphics[scale=1]{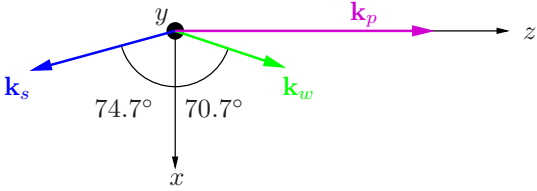}
	\caption{Orientation of the lasers' wave vectors for a Doppler-free three-photon transition.}
	\label{fig:laserangles}
\end{figure}

Doppler broadening is the main obstacle for LWI at short wavelengths. This becomes increasingly problematic if drive and probe field's frequencies differ strongly \citep{Lukin1996}, as is the case in the discussed scheme. To overcome this problem, \cite{Fry1999} proposes a Doppler-free three-photon transition. A particle moving at the velocity $\vec{v}$ relative to the emitter of the electromagnetic wave with angular frequency $\omega$ and wave vector $\vec{k}$ experiences a linear Doppler shifted frequency $\omega' = \omega - \vec{k} \cdot \vec{v}$. Therefore, a moving atom senses the Doppler-shifted three-photon detuning
\begin{equation}
\Delta_3 = \Delta_p + \Delta_s -\Delta_w - \left( \vec{k}_p + \vec{k}_s - \vec{k}_w \right) \cdot \vec{v}.
\end{equation}
If the fields satisfy the three-photon resonance for an atom at rest, i.\,e., $\Delta_p + \Delta_s -\Delta_w=0$, then this is true for every velocity group by choosing $\vec{k_p}+\vec{k}_s-\vec{k}_w = 0$, as shown in Fig.~\ref{fig:laserangles}.

The resulting velocity averaged linear susceptibility $\chi^{(1)} (T) = \int \chi^{(1)} ( \vec{v} ) f(\vec{v},T) d^3\vec{v}$ is defined by the Doppler distribution
\begin{align}
f(\vec{v},T) &= \left( \frac{m}{2 \pi k_B T} \right)^{3/2} \exp \left( -\frac{m \vec{v}^2}{2 k_B T}\right).
\end{align}
at the temperature $T$ with the atomic mass of mercury $m$ and the Boltzmann constant $k_B$.

\begin{figure}
	\centering
	\includegraphics[scale=1]{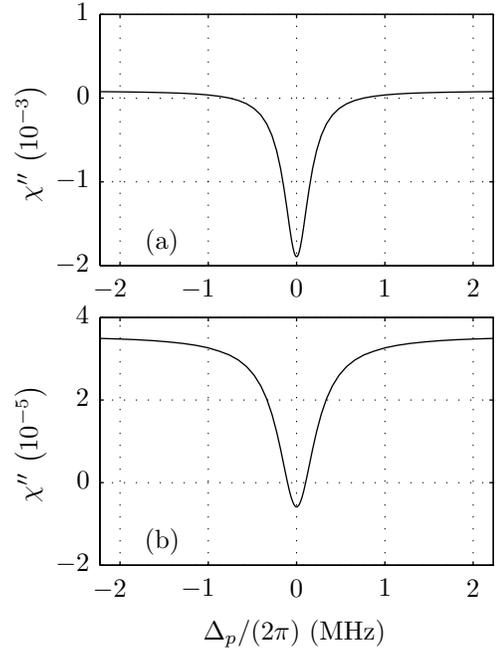}
	\caption{Velocity averaged absorption spectra $\chi''(\Delta_p,T)$ of the 13-level system versus probe field detuning $\Delta_p$. The spectrum shown in (a) was calculated for atoms at rest, while (b) is the Doppler average at $T=300$ K, $\Delta_w = \Delta_s = \Delta_r = 0$, $\Omega_s^{\pm1} = 2\pi \times 33.5$ MHz, $\Omega_w^{\pm1} = 2\pi \times 3.7$ MHz, $\Omega_r^{\pm1} = 2\pi \times 2.8$ MHz, $r = 1.1$ MHz, and $r_{cd} = 10$ MHz.}
	\label{fig:doppler13ls}
\end{figure}

In Fig.~\ref{fig:doppler13ls} (a), the central narrow feature of LWI for atoms at rest is shown (cf. Fig.~\ref{fig:fourls}). At finite temperature, this gain peak persists even though at reduced magnitude, see Fig.~\ref{fig:doppler13ls} (b). Both spectra are approximately Lorentzian with a width (FWHM) of $171$ kHz (a) and $256$ kHz (b). As a comparison, the probe transition's Doppler width (FWHM) for a temperature of $300$ K is $1.04$ GHz.


\section{Other broadening mechanisms}
\label{otherBroad}
In this section, we discuss further broadening mechanisms that may affect the performance of LWI and estimate their influence on the considered system for realistic experimental conditions.

For the atomic density and temperature under consideration (cf. Sec.~\ref{radDamp}), we estimate the ratio of dephasing rate to radiative coherence decay rate \cite{Lewis1980, Beyer2009, Laporte1979, OBrien2011} to $\gamma_{ab}^\text{deph}/\gamma_{ab}^\text{rad} = 0.07$, $\gamma_{ca}^\text{deph}/\gamma_{ca}^\text{rad} = 0.08$, and $\gamma_{cd}^\text{deph}/\gamma_{cd}^\text{rad} = 0.06$. The collisional dephasing rates of all other coherences are considerably smaller than these \cite{Beyer2009}. Therefore, we neglect collisional line broadening in our calculations. The coherence $\rho_{bd}$ is of special importance since it is closely connected to the aforementioned three-photon transition. We neglect its dephasing rate $\gamma_{bd}^\text{deph}$ assuming that it is considerably smaller than the coherence decay induced by the technical noise of the driving lasers we elaborate on later.

Further, we estimate the frequency shift induced by the recoil an atom exhibits during absorbing or emitting a photon \cite{Guo1992,Horak1995}. For the lasing transition this shift is $\delta \omega_p^\text{rec} = \hbar \omega_p^2/(2 m c^2) = 2 \pi \times 15$ kHz, with the atomic mass of mercury $m=200.6$ u and the speed of light $c$. For the other relevant transitions this shift is even smaller. The results of realistic calculations of the gain peak, presented later in this paper, show a width more than an order of magnitude larger than this shift. Therefore, we neglect the effect of recoil shifts in our calculations. However, it is worth mentioning that for possible generalizations of the discussed scheme towards shorter wavelength recoil effects may need to be considered.


\section{Technical noise of driving fields}
\label{noise}
To implement the driving laser fields $\vec{E}_s$ and $\vec{E}_w$, we plan to utilize single mode external cavity diode lasers, which exhibit predominately phase diffusion \cite{Haslwanter1988, Osinski1987} with linewidths below $1$ MHz. Further, the repump field $\vec{E}_r$ will be implemented as a spectrally broadened laser with a linewidth of $\sim 25$ MHz. Clearly, technical fluctuations in the phase affect coherent multi-photon processes negatively. Therefore, we will investigate the influence of phase noise on the LWI system's gain in this section. The influence of phase noise on the absorption spectra has been discussed for two-level atoms \cite{Haslwanter1988, Ritsch1990, Walser1994} and three-level LWI schemes \cite{Fleischhauer1994, Gong1995, Vemuri1994, Sultana1994}. A lucid introduction to stochastic methods can be found in \cite{Gardiner2003,Kampen1981}.

We start by writing the complex amplitudes of the slowly varying driving fields (cf. \eq{eqn:electricfields}) as
\begin{equation}
	\mathcal{E}_j (\vec{r},t) = \mathcal{E}_j(\vec{r}) e^{i \varphi_j(t)},
\end{equation}
with the deterministic amplitude $\mathcal{E}_j(\vec{r})$, the stochastic phase $\varphi_j(t)$, and $j=s,w,r$. 

The phases are modeled as stochastic processes undergoing diffusion characterized by a linewidth $b_j$ satisfying the stochastic Ito differential equations
\begin{equation}
	d \varphi_j (t) = \sqrt{2 b_j} dW_j(t),
	\label{eqn:sdephi}
\end{equation}
with $W_j(t)$ being a standard Wiener process \cite{Gardiner2003} with a vanishing mean $\langle dW_j(t)\rangle =0$ and variance $\langle dW_j^2(t)\rangle=dt$. Therefore, the probability density function $p(\varphi_s, \varphi_w,\varphi_r, t)$ satisfies the pure diffusion Fokker-Planck equation
\begin{equation}
\partial_t p = ( b_s \partial_{\varphi_s}^2 + b_w \partial_{\varphi_w}^2 + b_r \partial_{\varphi_r}^2) p.
\end{equation}
The stochastic average of a quantity $X(\varphi_s, \varphi_w,\varphi_r, t)$ with respect to the stochastic phases is given by
\begin{equation}
\langle X \rangle =  \iiint_0^{2\pi}  p \, X \,d\varphi_s d\varphi_w  d\varphi_r.
\end{equation}
It is worth to note the difference between this stochastic average, the quantum mechanical average, and the Doppler average. Throughout this paper $\langle \cdot \rangle$ will always denote the stochastic average over the phase fluctuations. This stochastic phase-diffusion model (PDM) leads to a stationary field autocorrelation function
\begin{equation}
\langle \mathcal{E}_j^*(\vec{r},t+\tau)\mathcal{E}_j(\vec{r},t) \rangle=|\mathcal{E}_j(\vec{r})|^2 e^{-b_j|\tau|},
\end{equation}
and a Lorentzian spectrum with linewidth $b_j$.

The Bloch equations depend functionally on the stochastic driving fields, therefore they become stochastic differential equations as well. Applying the rules of stochastic Ito calculus, one can derive ordinary differential equations for the stochastic averaged Bloch equations (cf. Appendix \ref{appendixPDM}).

The physical observables are stochastic averaged populations $\langle \rho_{ii} \rangle$ and contributions to the dipole energy $\langle \rho_{ij} e^{i \varphi_k (t)} \rangle$. Therefore, it is useful to introduce a unitary transformation of the stochastic density operator
\begin{equation}
\hat{\varrho}(t) = \hat{U}(t) \hat{\rho}(t) \hat{U}^\dagger(t), \label{eqn:transformation}
\end{equation}
that gauges away the stochastic phases. The resulting equations of motion for the averaged density operator $\langle \hat{\varrho} \rangle$ over the driving fields' phase fluctuations are given by
\begin{equation}
	\partial_t \langle \hat{\varrho} \rangle = ( \mathcal{L}_c + \mathcal{L}_i + \mathcal{L}_{pd} ) \langle \hat{\varrho} \rangle.
	\label{eqn:PDMMaster}
\end{equation}
The full expression of the phase diffusion Liouvillian $\mathcal{L}_{pd}$ can be found in the Appendix \ref{appendixPDM}. It contains additional damping terms for the atomic coherences depending on the fields' linewidths.

\begin{figure}
	\includegraphics[scale=1]{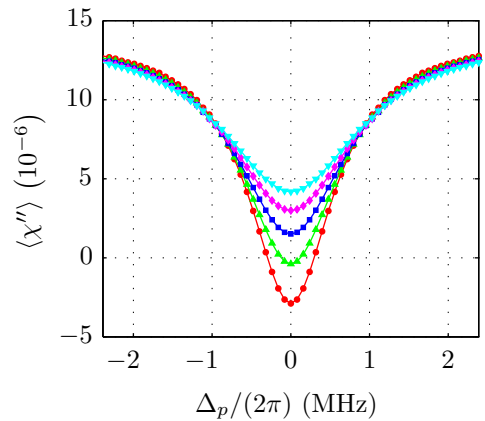}
	\caption{Velocity averaged, phase noise averaged absorption spectra $\langle \chi'' (\Delta_p,T)\rangle $ versus detuning $\Delta_p$ on the probe transition, for different linewidths of the driving fields $b_s=b_w= 0$ kHz (red dots), $b_s=b_w= 2\pi\times 8$ kHz (green triangles), $b_s=b_w= 2\pi\times 16$ kHz (blue squares), $b_s=b_w= 2\pi\times 24$ kHz (magenta diamonds), $b_s=b_w= 2\pi\times 32$ kHz (cyan triangles), and $b_r=2\pi \times 25$ MHz for all lines. All other parameters are chosen as in Fig.~\ref{fig:doppler13ls}.}
	\label{fig:3dnoise}
\end{figure}

In Fig.~\ref{fig:3dnoise} the resulting Doppler averaged absorption spectra calculated from \eq{eqn:PDMMaster} are shown for different linewidths.
The gain peak of the calculated absorption spectra shows a strong dependency on the driving fields' linewidths. While its width increases with increasing linewidths, its magnitude decreases. 

\begin{figure}
\centering
\includegraphics[scale=1]{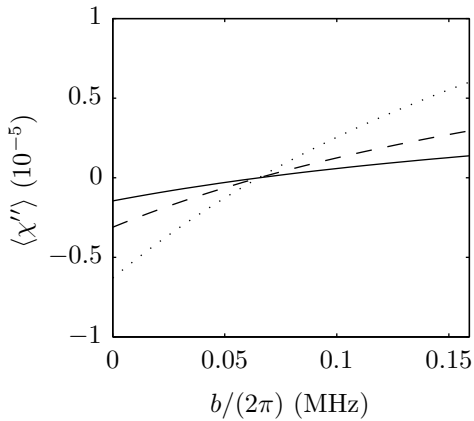}
\caption{Velocity averaged, phase noise averaged absorption $\langle \chi'' \rangle$ is plotted for temperatures of $290$ K (solid line), $300$ K (dashed line), and $310$ K (dotted line) against the driving fields' linewidth $b_s=b_w=b$. $\Delta_p=0$ whereas all other parameters are chosen as in Fig.~\ref{fig:doppler13ls}.}
\label{fig:linewidthDependence}
\end{figure}

The dependency of the susceptibility's imaginary part at resonance $\Delta_p = 0$ on the driving fields' linewidths is shown in Fig.~\ref{fig:linewidthDependence} for different temperatures. For increasing temperatures in the vapor cell the atomic density and with it the linear gain increases. However, this comes at the cost of increasing collisional dephasing rates that affect the atomic coherence used for LWI. By neglecting these rates we are not able to determine the optimal density at which the trade-off between  optical depth and collisional dephasing maximizes the gain. This has to be determined experimentally.

\begin{figure}
\centering
\includegraphics[scale=1]{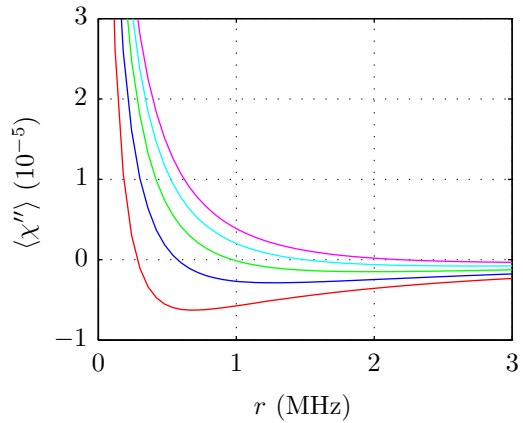}
\caption{Velocity averaged, phase noise averaged absorption $\langle \chi'' \rangle$ plotted against the incoherent pump rate $r$ on the probe transition for $b= 0$ kHz (red line), $b= 2\pi\times 16$ kHz (blue line), $b= 2\pi\times 32$ kHz (green line), $b= 2\pi\times 48$ kHz (cyan line), $b= 2\pi\times 64$ kHz (magenta line). All other parameters are chosen as in Fig.~\ref{fig:doppler13ls}.}
\label{fig:pumpDependence}
\end{figure}

Fig.~\ref{fig:pumpDependence} shows the dependency of the linear gain on the strength of the incoherent pump applied to the probe transition for different linewidths. This result shows that the loss in linear gain caused by phase noise can to some extent be compensated by stronger pumping on the probe transition.


\section{Expected laser power}
\label{lasertheory}
In the previous sections the gain medium's linear response to the external probe field was investigated, whereas in this section the field $\vec{E}_p$ will be treated as a dynamical quantity, the lasing field. By applying semiclassical laser theory \cite{Lamb1964,Sargent1974,Haken1985,Siegman1986} this will lead to the stationary laser power.

The dynamics of the lasing field is determined by the Maxwell equations from which the wave equation
\begin{equation}
\left( \frac{1}{c^2} \partial_t^2 + \mu_0 \sigma \partial_t - \Delta \right) \vec{E}_p (\vec{r},t) = - \mu_0 \partial_t^2 \vec{P}_p (\vec{r},t),
\label{eqn:waveeqn}
\end{equation}
can be deduced under the assumptions that the charge density, the gradient of $\vec{P}_p$, and the magnetization vanish and the current density is given by conductivity $\sigma$ times the electric field $\vec{E}_p$. The polarization density $\vec{P}_p$ couples this wave equation to the medium's Bloch equations.

To solve this problem, we start by expanding the laser field
\begin{align}
\mathcal{E}_p (\vec{r},t) &= \sum\limits_n \mathcal{E}_n (t) u_n (\vec{r}) \exp \left[ i \phi_n (t) \right],\label{eqn:expEA}
\end{align}
in the resonator's Hermite-Gauss modes $u_n(\vec{r})$ (cf. Appendix \ref{appendixHermiteGauss}) that are known to accurately describe modes in ordinary optical resonators \cite{Siegman1970}. $\phi_n$ is the phase and $\mathcal{E}_n$ is the real amplitude of the $n$-th mode. Choosing a convenient normalization, the Hermite-Gauss modes obey the orthogonality relation
\begin{equation}
\int_{\mathbb{R}^3} u_m^* (\vec{r})   u_n (\vec{r}) d^3\vec{r} = V_c \delta_{mn},
\label{eqn:ortho}
\end{equation}
with $V_c$ being the resonator's mode volume. The next step is to project \eq{eqn:waveeqn} onto $u_n$ by using Eqs. \eqref{eqn:electricfields}, \eqref{eqn:expEA}, and \eqref{eqn:ortho}. If we assume the paraxial approximation to be valid, small losses in the resonator and further apply the slowly varying envelope approximation (SVEA), we obtain the equation of motion for the $n$th mode amplitude
\begin{align}
\partial_t \mathcal{E}_n (t) + \frac{\sigma}{2\epsilon_0} \mathcal{E}_n (t) = - \frac{\omega_p V_0^{(n)}}{2 \epsilon_0 V_c} \Im \left[ \mathcal{P}_n (t) \right], \label{eqn:eamplitude}
\end{align}
with
\begin{align}
V_0^{(n)} &= \int_\mathcal{V} u_n^* (\vec{r})   u_n (\vec{r}) d^3\vec{r},\\
\mathcal{P}_n(t) &= \int_\mathcal{V} u_n^* (\vec{r})   \mathcal{P}_p(\vec{r},t) e^{-i\phi_n(t)} d^3\vec{r},
\end{align}
and $\mathcal{V}$ denoting the volume of the gain medium. In the derivation of \eq{eqn:eamplitude}, we used the dispersion relation $k_p=\omega_p/c$ for the wave vector of the Hermite-Gauss modes and the fact that these modes approximately obey the Helmholtz equation (cf. Appendix \ref{appendixHermiteGauss}). In the following we will assume that the amplitude $\mathcal{E}_0$ of the TEM$_{00}$ is much larger than all higher amplitudes $\mathcal{E}_n$ with $n>0$. Consequently, these higher amplitudes will be neglected. For convenience we drop the index $0$.

In order to solve \eq{eqn:eamplitude}, the dependency of the medium's polarization amplitude $\mathcal{P}$ on the amplitude of the laser field $\mathcal{E}$ needs to be known. This dependency is determined by the Bloch equations averaged over the Maxwell-Boltzmann distribution and the phase fluctuations exhibited by the external fields. As we are interested in the stationary limit, we expand the medium's polarization density in powers of the lasing field with the instantaneous (non-)linear susceptibilities $\langle \chi^{(n)} \rangle$ using the ansatz
\begin{equation}
\vec{P}_p^{(+)} = \epsilon_0 \sum\limits_{m=0}^\infty \langle \chi^{(2m+1)} \rangle \left(\vec{E}_p^{(+)} \cdot \vec{E}_p^{(-)}\right)^{m} \vec{E}_p^{(+)}.
\label{eqn:expE}
\end{equation}
These susceptibilities are averaged over the driving fields' phase fluctuations and the Maxwell-Boltzmann velocity distribution and are assumed to be homogeneous over the laser modes spatial extension in the medium. This assumption will be rectified in the next section. It is worth noting that for \eq{eqn:expE} we assumed frequency conversion effects to be negligible. Furthermore, the usage of instantaneous susceptibilities causes \eq{eqn:expE} to be only valid if $\mathcal{E}$ varies slowly compared to the time it takes for the medium to respond. Using the expansion given in \eq{eqn:expE}, the polarization amplitude can be written as
\begin{align}
\mathcal{P} &= \frac{\epsilon_0}{V_0} \sum\limits_{m=0}^\infty \langle \chi^{(2m+1)} \rangle \ V_m \mathcal{E}^{2m+1},\label{eqn:proChiexp}\\
&= \mathcal{N} d_{ab} \sum\limits_{m=0}^\infty \frac{V_m}{(2m+1)!}\frac{\partial^{2m+1} \tilde{\rho}_{ab}}{\partial \Omega^{2m+1}} \bigg|_{\Omega=0} \Omega^{2m+1},\label{eqn:proRhoexp}
\end{align}
with $\Omega = d_{ab} \mathcal{E}/\hbar$ and
\begin{align}
V_m &= \int_\mathcal{V}\left\vert u(\vec{r}) \right\vert^{2m+2} d^3\vec{r},\\
\tilde{\rho}_{ab} &= \int_\mathcal{V} u^* (\vec{r})   \langle \rho_{ab}(\vec{r}) \rangle e^{-i\phi} d^3\vec{r}.
\end{align}
We calculate the expansion parameter of \eq{eqn:proRhoexp} by fitting a polynomial of degree $2M+1$ to the numerical calculated $\tilde{\rho}_{ab}(\Omega)$. A coefficient comparison between Eqs. \eqref{eqn:proChiexp} and \eqref{eqn:proRhoexp} then yields the nonlinear susceptibilities. Our calculations show a truncation with $M=2$ to be sufficient. We introduce the photon number as $n=2 \epsilon_0 V_c \mathcal{E}^2/ (\hbar \omega_p)$. Inserting \eq{eqn:proChiexp} in \eq{eqn:eamplitude} and truncating the series at $m=2$, we arrive at
\begin{equation}
\partial_t n = \alpha n - \beta n^2 - \gamma n^3,
\end{equation}
with the linear gain parameter
\begin{align}
\alpha &= -\frac{\sigma}{\epsilon_0} - \frac{\omega_p V_0}{V_c} \chi_1'',
\end{align}
the nonlinear saturation parameters
\begin{align}
\beta &= \frac{\hbar \omega_p^2 V_1}{2 \epsilon_0 V_c^2} \chi_3''&
\gamma &= \frac{\hbar^2 \omega_p^3 V_2}{4 \epsilon_0 V_c^3} \chi_5'',
\end{align}
and $\chi_m'' = \Im (\langle \chi^{(m)}\rangle)$. Thus, the stationary photon number is given
\begin{equation}
n_\text{st} =\left\{\begin{array}{cl} - \frac{\beta}{2\gamma} + \sqrt{\frac{\beta^2}{4\gamma^2}+\frac{\alpha}{\gamma}} = \frac{\alpha}{\beta} + O(\gamma), & \mbox{ }\alpha>0\\ 0, & \mbox{ }\alpha \le 0 \end{array}\right. .
\end{equation}
The sign of the linear gain parameter $\alpha$ indicates if the laser system is above ($\alpha>0$) or below ($\alpha \le 0$) threshold, while $\beta$ and $\gamma$ are saturation parameters that determine the stationary power when above threshold. For $\gamma \to 0$, we obtain the standard form of the photon number equation for a laser \cite{Haken1985}.

\begin{figure}
	\centering
	\includegraphics[scale=1]{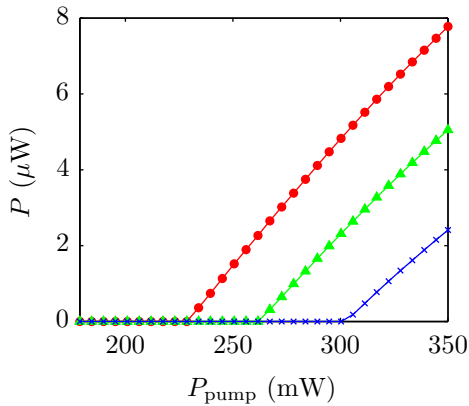}
	\caption{Stationary laser power $P$ inside the cavity versus pumping power $P_\text{pump}$ for $b_s = 2\pi\times 45$ kHz and $b_w= 2\pi\times 21.6$ kHz (red dots), $b_s = 2\pi\times 50$ kHz and $b_w= 2\pi\times 24$ kHz (green triangles), $b_s = 2\pi\times 55$ kHz and $b_w= 2\pi\times 26.4$ kHz (blue crosses). The parameters $T=300$ K, $A=4$ mm$^2$, $Q= \epsilon_0 \omega_p/\sigma=198 \times 10^6$ (quality factor), and $V_0/V_c=0.01$ were used. All other parameters are chosen as in Fig.~\ref{fig:doppler13ls}.}
	\label{fig:power}
\end{figure}

Assuming the laser mode to be transversely well localized in the medium, the approximation $V_m = V_0/2^m$ is justified. Further, the stationary laser power can be calculated to $P=\hbar \omega_p c \pi w_0^2 n/(2 V_c)$ with $w_0$ being the mode's beam waist. Fig.~\ref{fig:power} shows the dependency of the calculated stationary laser power on the pumping power $P_\text{pump}$ for different linewidths of the driving fields. The pumping power is related to the incoherent pump rate $r$ by
\begin{equation}
P_\text{pump} = \frac{\sqrt{2}\hbar \omega_p^3 \sigma_\omega A}{\sqrt{\pi^3} c^2 \Gamma_{ab}} r
\end{equation}
We assume the pump field to have a Gaussian spectrum with central frequency $\omega_p$ and variance $\sigma_\omega = 2\pi \times 440$ MHz corresponding to a Doppler spectrum at temperature $T=300$ K. Further, $A$ denotes the effective area cross-section of the pump field's beam. The graphs show the expected behavior, no lasing until a threshold pump power $P_\text{thr}$ and a merely linear dependency on the pump power above threshold. Increasing linewidths shift the threshold towards larger pump intensities and result in a flatter slope.

\begin{figure}
	\centering
	\includegraphics[scale=1]{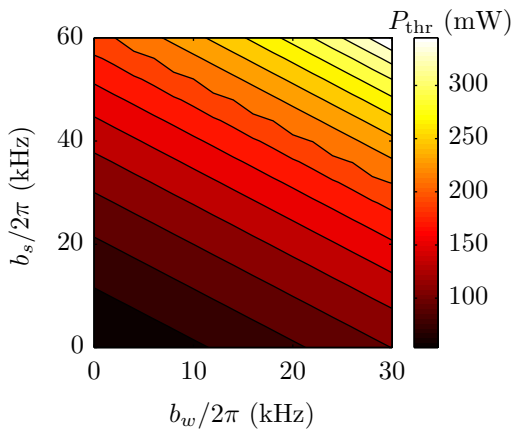}
	\caption{Threshold pumping power $P_\text{thr}$ is plotted against the external laser linewidths $b_s$ and $b_w$. The same parameters as in Fig.~\ref{fig:power} were used.}
	\label{fig:thresholdPower}
\end{figure}

Fig.~\ref{fig:thresholdPower} shows the dependency of the threshold pump power on the driving fields' linewidths. This graph reveals the importance of reducing the phase noise of the external laser for the feasibility of this experiment.


\section{Resonator modes}
\label{resonatorModes}
In the previous section, it has been assumed that the spatial distribution of the laser field is that of a Gaussian mode. In this section this assumption's validity will be proved by calculating the spatial gain distribution in the medium. Using this result, we obtain the real modes of the laser system. Assuming the laser field $\vec{E}_p$ to be highly monochromatic, we can neglect the time dependency of its amplitude $\mathcal{E}_p$ and write
\begin{equation}
\mathcal{E}_p (\vec{r}) = \psi(x,y;z) \exp(i \vec{k}_p \cdot \vec{r}).
\end{equation}
In the paraxial approximation, we assume that $x$ and $y$ are variables and $z$ becomes the chronological parameter. Within the linear response approximation the slowly varying envelope $\psi(z) \equiv \psi(x,y;z)$ obeys the Schr\"odinger-like wave equation
\begin{equation}
i \partial_z \psi (z) = [T+V(z)] \psi (z).
\label{eqn:waveeqnparax}
\end{equation}
We have used the following definitions
\begin{align}
T \psi(z) &\equiv -\frac{1}{2 k_p} (\partial_x^2 + \partial_y^2) \psi(x,y;z), \\
V(z) \psi(z)  &\equiv -\frac{k_p}{2} \langle \chi^{(1)}(x,y;z) \rangle \psi(x,y;z),
\end{align}
and the wave vectors norm $k_p = \vert \vec{k}_p \vert$. It is worth noting that $V(z)$ is a non-Hermitian complex potential accounting for changes in the index of refraction and absorption.
\subsection{Propagation through the empty resonator}
Before including the gain medium in our considerations, we will calculate the modes of the empty open resonator. In the following, we will shortly review the basic procedure of light propagation in open cavities \cite{Born1959, Siegman1986, Hodgson2004}.

In free space ($\chi^{(1)}=0$) the solution to \eq{eqn:waveeqnparax} is given by
\begin{equation}
\psi(z) = \mathcal{F}(z-z') \psi(z') = e^{-i T (z-z')} \psi(z'),
\label{eqn:fresnelIntegral}
\end{equation}
with the initial field distribution $\psi(z')$. The explicit form of \eq{eqn:fresnelIntegral} for $\psi(x,y;z)$ is just the Fresnel diffraction integral and can be evaluated efficiently with fast Fourier transform algorithms.

To simulate the round-trip in an optical resonator the action of a mirror with radii of curvature $R_x$ and $R_y$ and aperture function $\mathcal{A}(x,y)$ on the field $\psi$ needs to be specified
\begin{equation}
\mathcal{M} \psi(z) \equiv e^{-ik_p ( x^2/R_x + y^2/R_y )} \mathcal{A}(x,y) \psi(x,y;z).
\end{equation}

\begin{figure}
\includegraphics[scale=1]{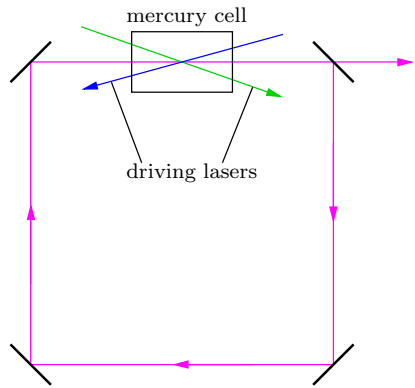}
\caption{Scheme of the LWI system's resonator. The one-directional ring resonator design and the orientation of the driving fields, represented by the blue and the green arrow respectively, is enforced by the Doppler free configuration described earlier.}
\label{fig:cavity}
\end{figure}

Combining free propagation with equal length $L$ and the action of the four mirrors in the ring cavity shown in Fig.~\ref{fig:cavity}, defines a round trip operator $\mathcal{R} \equiv [\mathcal{M} \mathcal{F}(L)]^4$. This yields an eigenvalue problem
\begin{equation}
\mathcal{R} \psi_\gamma = \gamma \psi_\gamma,
\end{equation}
for the eigenmodes $\psi_\gamma$ labeled by the complex eigenvalue $\gamma$. The mirrors have a curvature
radius $R$ and a tilt angle of $45^\circ$. We choose the $x$-direction to be in the sagittal plane and the $y$-direction to be in the tangential plane. From this follows $R_x = R \cos(45^\circ)$ and $R_y = R / \cos(45^\circ)$.

\begin{figure}
	\includegraphics[scale=1]{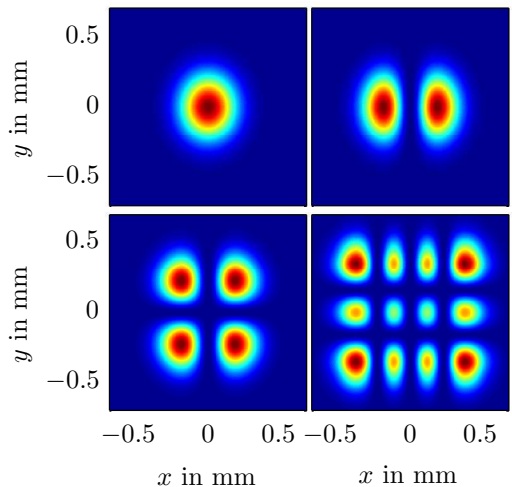}
	\caption{Transverse intensity patterns on the output mirror are shown for the modes TEM$_{00}$, TEM$_{10}$, TEM$_{11}$, and TEM$_{32}$ of the empty ring resonator. A quadratic aperture with a diameter of $1.38$ mm was used together with mirror radii $R=100$ cm and distance $L=20$ cm.}
	\label{fig:modes}
\end{figure}

Starting from Fox and Li \cite{Fox1960}, different methods \cite{Siegman1970, Southwell1981, Yuanying2004, New2012} have been developed to solve this eigenvalue problem. In our work, we use the Arnoldi-Krylov method \citep{Latham1980,Golub1996} to find the largest eigenvalues $\gamma_{mn}$ and the corresponding eigenvectors $\psi_{mn}$, which are the self-consistent transverse electromagnetic modes (TEM) of the ring cavity defined by $\mathcal{R}$. Fig.~\ref{fig:modes} shows the calculated modes's intensity patterns on the output mirror. As expected, these modes show strong resemblance to the Hermite-Gauss functions used for the mode expansion in the previous section.
\subsection{Propagation in inhomogeneous media}
\label{geometry}
\begin{figure}
	\includegraphics[scale=1]{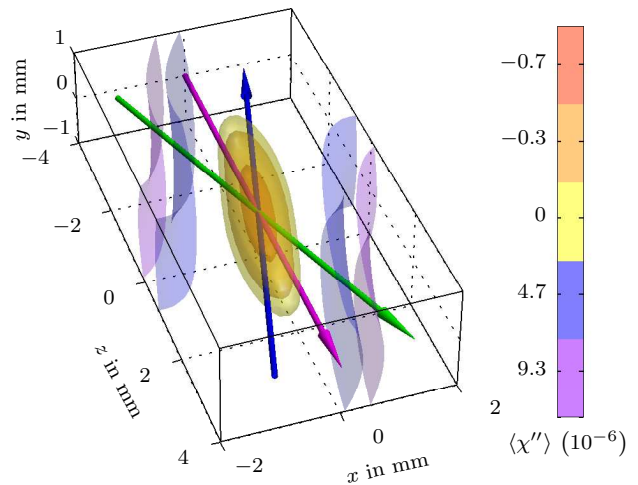}
	\caption{Iso-surfaces of the gain distribution $\langle \chi'' (x,y;z)\rangle$ versus position. The  green (weak driving laser), blue (strong driving laser) and violet (probe laser) arrows correspond to the laser beams' directions obeying the Doppler-free scheme. The external driving fields are Gaussian beams with waists $w_0 = 2$ mm, powers $P_s=200$ mW and $P_w=1.4$ mW, linewidths $b_s=2\pi \times 50$ kHz and $b_w=2\pi \times 24$ kHz and polarizations as given in \eq{eqn:pol}. The given laser powers correspond to peak Rabi frequencies of $\Omega_s^{\pm 1} = 2 \pi \times 33.5$ MHz and $\Omega_w^{\pm 1} = 2 \pi \times 3.7$ MHz respectively. $r=2.3$ MHz and all other parameters are chosen as in Fig.~\ref{fig:doppler13ls}.}
	\label{fig:spatialdistr}
\end{figure}

In Sec.~\ref{lasertheory}, we have assumed spatially homogeneous susceptibilities to calculate the stationary intensity of the laser field. This does not correspond to the real situation where the intersection of external Gaussian laser beams creates a tilted, ellipsoidal gain distribution as shown in Fig.~\ref{fig:spatialdistr}. The orientations of the driving fields' wave vectors are fixed by the three-photon Doppler free configuration in Fig.~\ref{fig:laserangles}.

The central yellow/orange structure provides gain for the laser while in the blue structures on both sides amplified absorption is observed. This can be explained by considering the ratio $\Omega_s/\Omega_w$ in both areas. The gain region is basically centered around the laser beam associated with $\Omega_s$ resulting in the mentioned ratio to be large, while increased absorption is observed in regions with a small ratio $\Omega_s/\Omega_w$. For small $\Omega_s$ the described coherence effect is not observed since the needed Autler-Townes splitting cannot be achieved whereas a relatively large $\Omega_w$ prevents population trapping in the state $\ket{d}$ and by this means increases absorption on the lasing transition.

To investigate the influence of the inhomogeneous gain distribution $\langle \chi^{(1)}(\vec{r}) \rangle$ on the mode structure, we need to propagate the field through the medium with length $L_m$
\begin{equation}
\psi (z_f) = \mathcal{K} \psi(z_i) \equiv \mathcal{T} e^{-i\int_{z_i}^{z_f} [T+V(z)] dz} \psi(z_i),
\end{equation}
with $z_i = -L_m/2$ and $z_f = L_m/2$. This formal definition of the chronological ordered ($\mathcal{T}$) propagator $\mathcal{K}$ is evaluated approximately by splitting the medium into $N$ short slices of length $\delta z = L_m/N$ and using the split operator method \citep{Fleck1976,VanRoey1981}
\begin{equation}
\mathcal{K} = e^{-i T \frac{\delta z}{2} } \left(\prod\limits_{l=0}^{N-1} e^{ -i V(z_l) \delta z} e^{ -i T \delta z}\right)  e^{i T \frac{\delta z}{2} },
\end{equation}
where we have evaluated the potential at positions $z_l = z_i + l \delta_z$.

Now, this modifies the round trip operator of the cavity and the gain medium to
\begin{equation}
\mathcal{R}' = e^{-\nu} \mathcal{M} \mathcal{F}(L') \mathcal{K} \mathcal{F}(L') [\mathcal{M} \mathcal{F}(L)]^3
\end{equation}
with $L'=(L-L_m)/2$. The additional factor $e^{-\nu}$ phenomenologically models the cumulative loss per round trip caused by output coupling and imperfection of optical elements. The corresponding resonator modes can again be calculated using the Arnoldi-Krylov method.
\subsection{Results}
The crucial parameters for the experiment are the gain per round-trip $\vert \gamma \vert^2$ and the beam quality that can be measured by the $M^2$ parameter \cite{Siegman1993,ISO11146} defined for the transverse $x$-direction as
\begin{equation}
M^2_x = \frac{\pi \vartheta_x d_x}{4 \lambda},
\end{equation}
with the wavelength $\lambda$, the divergence angle in $x$-direction $\vartheta_x$, and beam diameter in $x$-direction $d_x$ at the beam's waist. Hermite-Gauss modes of order $m$ have a beam quality characterized by $M_x^2 = 2m +1$.

The most important control parameter for the gain medium's structure are the spatial field distributions of the driving lasers. We assume the driving fields to be Gaussian beams focussed in the center of the gain medium. Consequently, the field distributions are determined by the beams' waists. Since the driving fields' peak intensity is constrained to have optimal gain in the focal region, a certain beam waist corresponds to a certain power required.

\begin{figure}
	\centering
	\includegraphics[scale=1]{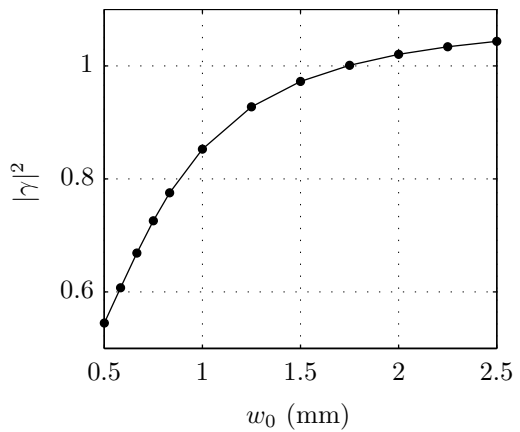}
	\caption{Linear amplification $|\gamma_{00}(w_0)|^2$ per cavity round-trip for TEM$_{00}$ plotted against the driving fields' waist $w_0$. The cavity design shown in Fig.~\ref{fig:cavity} was used with the same parameters as in Fig.~\ref{fig:modes} and $e^{-\nu}=0.95$ . All other parameters are chosen as in Fig.~\ref{fig:spatialdistr}.}
	\label{fig:roundTripAmpl}
\end{figure}

\begin{figure}
	\centering
	\includegraphics[scale=1]{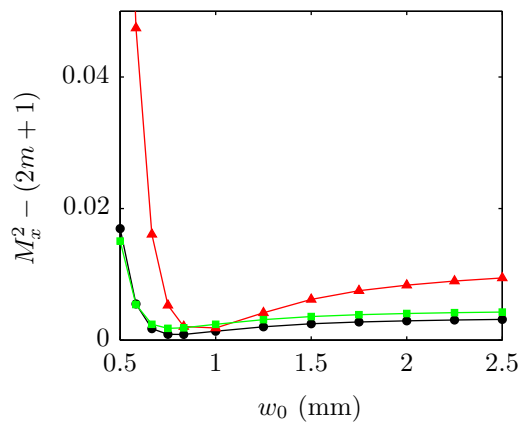}
	\caption{The beam quality parameter $M_x^2$ plotted against the driving fields' waists for TEM$_{00}$ (black dots), TEM$_{01}$ (green squares), and TEM$_{10}$ (red triangles). All parameters are chosen as in Fig.~\ref{fig:roundTripAmpl}.}
	\label{fig:msquare}
\end{figure}

Figs.~\ref{fig:roundTripAmpl} and \ref{fig:msquare} show the dependencies of the linear amplification per round-trip $\vert \gamma \vert^2$ and the beam quality parameter $M_x^2$ on the driving fields' waist $w_0$ respectively. For small $w_0$ the laser field is attenuated in the medium, resulting in $\vert \gamma \vert^2 <1$, whereas for larger $w_0$ the gain region increases and at some point the gain in the medium compensates for the absorption and diffraction losses ($\vert \gamma \vert^2=1$); for yet larger $w_0$ the field is amplified after each round-trip and the laser process starts. If $w_0$ is much larger than the laser field's extension in the gain medium, the gain saturates. In the case of large $w_0$ the beam quality of the laser field reaches the beam quality of the empty resonator modes (cf. Fig.~\ref{fig:modes}). For small $w_0$ the beam quality is worse, $M^2_x$ exceeds the former mentioned value considerably. This can be explained by interpreting the gain region in the medium as an effective aperture for the laser beam. If this aperture is much larger than the transverse extension of the laser beam's intensity distribution, the aperture has no significant influence on the laser beam and the resulting modes are those of the empty resonator. For smaller $w_0$ the effective aperture of the gain region becomes smaller and the laser beam is diffracted at this aperture, resulting in poorer beam quality and larger $M^2_x$ values. In the intermediate regime one observes an minimum in $M^2_x$, optimal mode quality. This is obtained when the gain structure in the medium and the respective mode's intensity distribution match best. Diffraction patterns in the outer regions of the intensity distributions are absorbed while the relevant part of the mode is not diffracted.



\section{Conclusions}
We have developed a realistic multi-level model for the UV lasing scheme in mercury vapor proposed in \cite{Fry1999}, including technical noise of the driving field, Doppler broadening, the spatial inhomogeneous structure of the gain medium, and self-consistent eigenmodes of a four-mirror ring cavity. This model was used to identify crucial experimental parameters such as the linewidths and the waists of the driving fields and to investigate the dependency of linear gain, stationary power, and mode quality of the laser system on these parameters. The results of this analysis demonstrate the parameter ranges and the expected performance of the laser system.


\section*{Acknowledgments}
We acknowledge support by the Deutsche Forschungsgemeinschaft (DFG) grant WA 1658/2-1.


\appendix

\section{Details of PDM calculation}
\label{appendixPDM}
The stochastic phases $\varphi_s$, $\varphi_w$, and $\varphi_r$ appear in the Bloch equations as parameters. To separate the stochastic and deterministic dynamics, we apply the transformation given in \eq{eqn:transformation} with
\begin{align}
\hat{U}(t) &= \exp \bigg( \tfrac{i}{2} \sum\limits_k \varphi_k(t) \sum\limits_j \xi_{kj} \hat{s}_{jj} \bigg),\\
\xi &= \bordermatrix{ & a & b & c & d & e \cr
s & -1 & -1 & 1 & 1 & 1\cr
w & 1 & 1 & 1 & -1 & 1\cr
r & 1 & 1 & 1 & 1 & -1}.
\end{align}
The sum over $j$ extends over the atomic states whereas $k \in \{ s,w,r\}$. Applying the rules of Ito calculus \cite{Gardiner2003} on the variable substitution $d\hat{\rho} \to d\hat{\varrho}$, we obtain
\begin{equation}
\begin{split}
d  \hat{\varrho}  = &\mathcal{L} \hat{\varrho} dt + \frac{1}{4} \sum\limits_{k,k'} d\varphi_k d\varphi_{k'} \sum\limits_{i,j} \xi_{kj} \xi_{ki} \hat{s}_{jj} \hat{\varrho} \hat{s}_{ii}\\
&- \frac{1}{8} \sum\limits_{k,k'} d\varphi_k d\varphi_{k'} \sum\limits_j \left( \hat{s}_{jj} \hat{\varrho} + \hat{\varrho} \hat{s}_{jj} \right),
\end{split}
\label{eqn:ito1}
\end{equation}
with $\mathcal{L}=\mathcal{L}_c+\mathcal{L}_i$. It is worth noting that $\mathcal{L}$ is deterministic as the transformation $\hat{U}$ separated the deterministic evolution from the stochastic fluctuations. Under the assumption that $\hat{\varrho}$ is non-anticipating $\langle \hat{\varrho} d\varphi_k \rangle = 0$, the average of \eq{eqn:ito1} over the phase fluctuations is given by
\begin{equation}
\langle d\hat{\varrho} \rangle = (\mathcal{L}_c + \mathcal{L}_i + \mathcal{L}_{pd}) \langle \hat{\varrho} \rangle dt
\label{eqn:ito2}
\end{equation}
The phase-diffusion Liouvillian appearing in \eq{eqn:PDMMaster} is thus given by
\begin{equation}
\begin{split}
\mathcal{L}_{pd} \langle \hat{\varrho} \rangle = &\sum\limits_k \frac{b_k}{2} \sum\limits_{i,j} \xi_{kj} \xi_{ki} \hat{s}_{jj} \langle \hat{\varrho} \rangle \hat{s}_{ii} \\
&- \sum\limits_k \frac{b_k}{4} \sum\limits_j \left( \hat{s}_{jj} \langle \hat{\varrho} \rangle + \langle \hat{\varrho} \rangle \hat{s}_{jj} \right).
\end{split}
\end{equation}
For \eq{eqn:ito2} the equations \eqref{eqn:sdephi} for the stochastic phases were used together with the properties of the independent Wiener increments $\langle dW_k dW_k' \rangle = \delta_{kk'} dt$.

\section{Hermite-Gauss modes}
\label{appendixHermiteGauss}
The Hermite-Gauss mode \cite{Siegman1986} of order $(mn)$ is given by
\begin{equation}
u_{mn}(x,y,z) = N(z) u_m(x,z) u_n(y,z) e^{-ik_pz+ i \Phi_{mn} (z)}
\end{equation}
with the one-dimensional mode function
\begin{equation}
u_m(x,z) = H_m\bigg(\frac{\sqrt{2}x}{w_1(z)}\bigg) \exp\bigg( \frac{-x^2}{w_1^2(z)}- \frac{i k_p x^2}{2 R_1(z)}\bigg),
\end{equation}
the normalization factor $N(z)$, the Guoy phase $\Phi_{mn}(z)$, the beam width $w_j(z)$, and the curvature radius $R_j(z)$. $H_m$ denotes the $m$-th Hermite polynomial. In the case of the ring resonator it is important to note that the $z$-direction is the direction of the optical axis tilted by the mirrors and that the $x$- and $y$-direction are the corresponding local transverse directions.

The Hermite-Gauss modes are solutions to the paraxial wave equation in vacuum (cf. \eq{eqn:waveeqnparax}). In the paraxial case they approximately obey the Helmholtz equation
\begin{equation}
\Delta u_{mn}(\vec{r}) \approx - k_p^2 u_{mn}(\vec{r}).
\end{equation}

\bibliographystyle{osajnl}

\end{document}